\documentclass[a4paper,11pt]{article}
\usepackage{pos}

\title{Status and outlook of quark flavour physics}

\usepackage[absolute,overlay]{textpos}

\author*[a]{J. Tobias Tsang}

\affiliation[a]{CERN, Theoretical Physics Department\\
  Street number, , Geneva, Switzerland}

\emailAdd{j.t.tsang@cern.ch}

\abstract{
In recent years there has been impressive progress in quark flavour physics, with current efforts tackling complicated quantities such as for example inclusive decays, decays to QCD-unstable final states and radiative decays. At the same time current lattice flavour physics results are receiving a lot of attention from outside the lattice community. This requires careful scrutiny, even for well explored quantities.

In this plenary contribution I review the status of flavour physics observables and, in the context of heavy flavours, suggest possible benchmark quantities that could allow to better compare intermediate results. These can in turn be used to investigate the origin of existing (and possible future) tensions.

\begin{textblock}{20}(15.0,1.70)
CERN-TH-2025-050\\
\end{textblock}%
}

\FullConference{The 41st International Symposium on Lattice Field Theory (LATTICE2024)\\
 28 July - 3 August 2024\\
Liverpool, UK\\}

\begin{document}
\maketitle

\section{Introduction}
This review mostly concentrates on quantities that can be combined with experimental inputs in order to determine elements of the Cabibbo-Kobayashi-Maskawa (CKM) matrix or which allow for direct comparison to experimental observables. In both cases the ultimate goal is to scrutinise the Standard Model in order to find hints of new physics and to constrain the parameter space any new physics model can occupy. 

Quark flavour physics is a wide field comprising the full spectrum from very mature calculations with fully controlled uncertainties to novel and exploratory first calculations whose computations have only just become feasible. Due to the vast size of the field it is naturally impossible to review every single calculation of the field. The following must hence be a selection and any work not discussed in detail here is not omitted due to lack of interest, but rather due to limited time and space. I refer the interested reader to the proceedings contributions of work I could not cover here, as well as to recent reviews of the  literature~\cite{Meinel:2024pip, Kaneko:2024teu, FlavourLatticeAveragingGroupFLAG:2024oxs,Tsang:2023nay}. 

Naturally different strands of lattice QCD overlap as many methodologies are applicable to different observables. In particular the calculation of inclusive decays uses many techniques related to inverse problems and is covered in Will Jay's plenary contribution to this conference~\cite{Jay:2025dzl}. Similarly the correct treatment of multi-hadron final states in weak decays combines traditional flavour methodologies with those of scattering. This type of processes was also covered in Felix Erben's plenary contribution at this conference~\cite{Erben:2025zph}.

The remainder of this document is structured as follows. In Sections \ref{sec:breadnbutter1} and \ref{sec:breadnbutter} I will review mature calculations of ``bread and butter'' observables which address all sources of uncertainties, with a particular focus on heavy quark flavour physics. In Sec.~\ref{sec:breadnbutter} I will highlight a number of cases, where tensions between the theory predictions persist and then suggest some possible benchmark quantities that could be used to compare simpler intermediate quantities at a lower level of complication in Sec.~\ref{sec:benchmarks}. In Sec.~\ref{sec:newresults} I will highlight some select novel calculations followed by methodological improvements in Sec.~\ref{sec:methods} before providing an outlook in Sec.~\ref{sec:outlook}.

\section{Heavy quark masses and decay constants \label{sec:breadnbutter1}}
The determination of pseudoscalar heavy-light decay constants and heavy quark masses are well trodden terrain with a wealth of available and mature results which are by-in-large in good agreement~\cite{FlavourLatticeAveragingGroupFLAG:2024oxs,  Tsang:2023nay}. However, these quantities remain of interest, not least because they serve as excellent benchmarks to test novel approaches. RQCD/ALPHA recently computed the leptonic decay constants $f_{D_{(s)}}$ and their ratio, finding good agreement with the literature~\cite{Kuberski:2024pms}. HPQCD presented a study of the hyperfine splittings and ratios of pseudoscalar, tensor and vector decay constants in the $D_{(s)}^{(*)}$ and $B_{(s)}^*$ systems~\cite{Miller:2025dez}. The CLQCD collaboration recently presented new results for the pseudoscalar, vector and tensor decay constants of the $D^{(*)}_{(s)}$, $\eta_c$ and $J/\Psi$ as well as the charm quark mass~\cite{CLQCD:2024yyn}. The RBC/UKQCD collaboration employed a massive non-perturbative renormalisation scheme to compute the charm quark mass, which again is in good agreement with the literature~\cite{DelDebbio:2024hca}.

\section{Heavy semileptonic form factors \label{sec:breadnbutter}}
Due to the wealth of existing and forthcoming experimental data, form factors describing the semi-leptonic decays of mesons containing at least one heavy (charm or bottom) quark have recently received much attention and are heavily sought after. Generically, semi-leptonic decays are parameterised as 
\begin{equation}
  \frac{\mathrm{d}\Gamma (I \to F l_1 l_2)}{\mathrm{d} q^2} = \sum_{X}\mathcal{K}_X f_X^{I\to F}(q^2)\,,
\end{equation}
where the $\mathcal{K}_X$ contain known perturbative and kinematic factors, $q_\mu = (p_I - p_F)_\mu$ is the 4-momentum transfer from initial ($I$) to final ($F$) state and $f_X(q^2)$ are the non-perturbative form factors which we aim to compute on the lattice. The index $X$ runs over the contributing form factors, e.g. $f_0$ and $f_+$ ($f$, $g$, $F_1$, $F_2$) for tree-level decays into pseudoscalar (vector) final states. In the case of rare (i.e. loop-induced) decays additional tensor form factors contribute to the standard model prediction, whilst they only appear beyond the standard model for tree-level decays. The leptons in the final state are $l_1 l_2 \equiv \ell \nu$ for tree-level decays and $l_1 l_2 = \ell^+ \ell^-$ for rare decays.  
\subsection{Status of the literature}
Compared to the decay constants and quark masses, the situation here is more complicated due to the wealth of different decays and the fact that these objects are functions of a kinematic variable rather than scalar objects. Hence typically fewer results are available for any given quantity.  Furthermore, the current status of published results displays a number of tensions between predictions which need to be addressed. In particular the latest FLAG review~\cite{FlavourLatticeAveragingGroupFLAG:2024oxs} applies a $\sqrt{\chi^2/\mathrm{dof}}$ inflation to their averages of $D \to \pi \ell \nu$, $D \to K \ell \nu$, $B\to \pi\ell\nu$, $B_s \to K\ell\nu$ and $B\to K \ell\ell$ form factors. Even though the form factors of $B \to D^*\ell \nu$ form factors are not $\chi^2$ inflated\footnote{since averages between $N_f=2+1$ and   $N_f=2+1+1$ are taken separately, the potential of required $\chi^2$-error inflations is reduced here.}, in the combination with experimental data for the extraction of $V_{cb}$ a $\chi^2$ inflation is needed for both the $N_f=2+1$ and the $N_f=2+1+1$ average. Whilst these $\sqrt{\chi^2/\mathrm{dof}}$ inflations are only indicative, as they are based on an interpretation of the total uncertainties as purely statistical, clearly the origin of these tensions needs to be understood -- in particular as long as ``we'' as a community effectively interpret systematic uncertainties in a statistical sense by adding uncertainties in quadrature.

\begin{figure}
    \centering
    \includegraphics[width=0.5\linewidth]{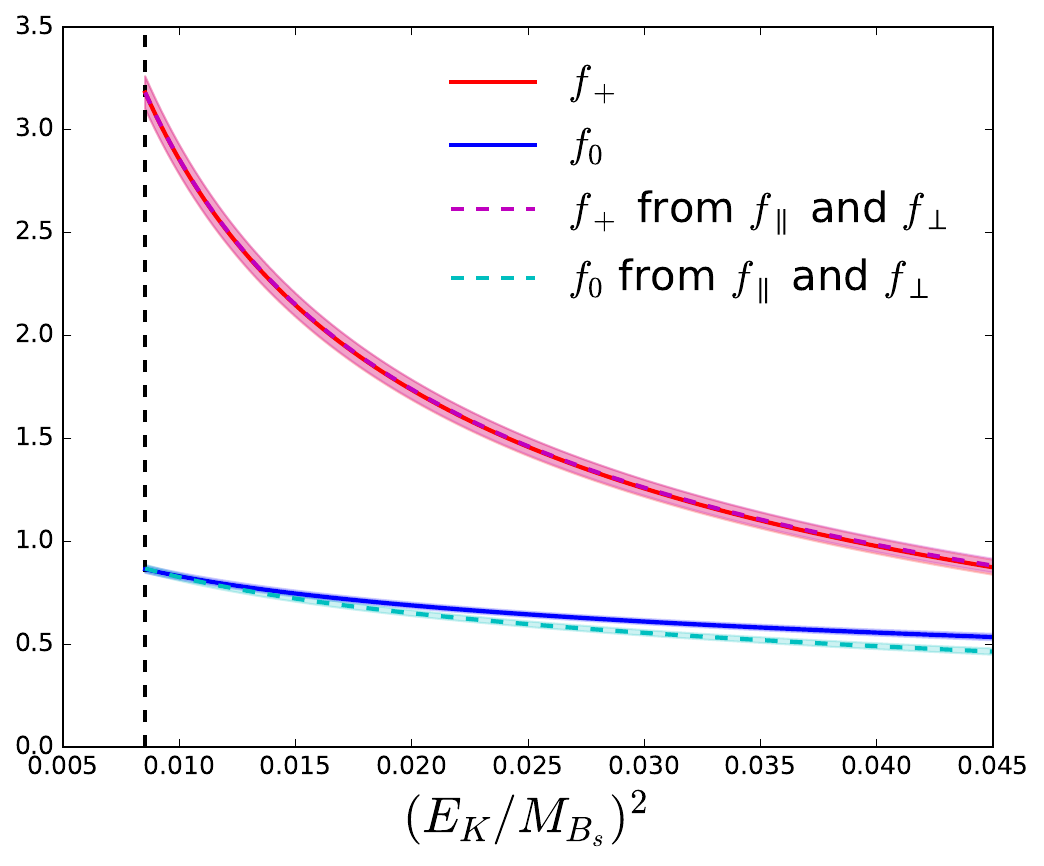}
    \includegraphics[width=0.45\linewidth]{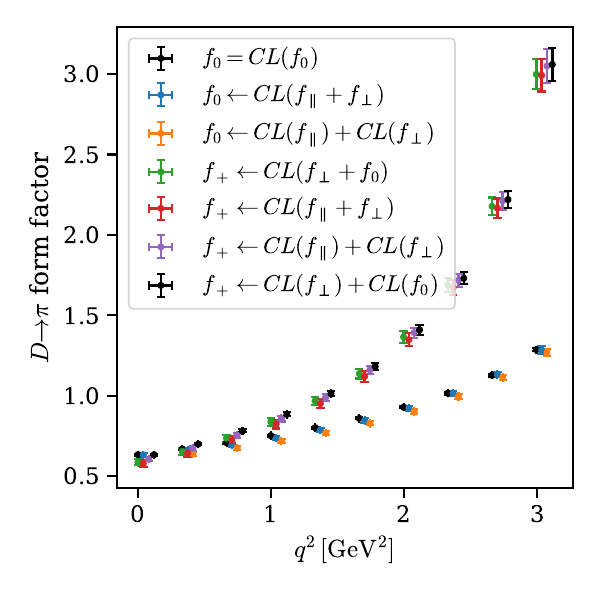}\\
    \includegraphics[width=0.5\linewidth]{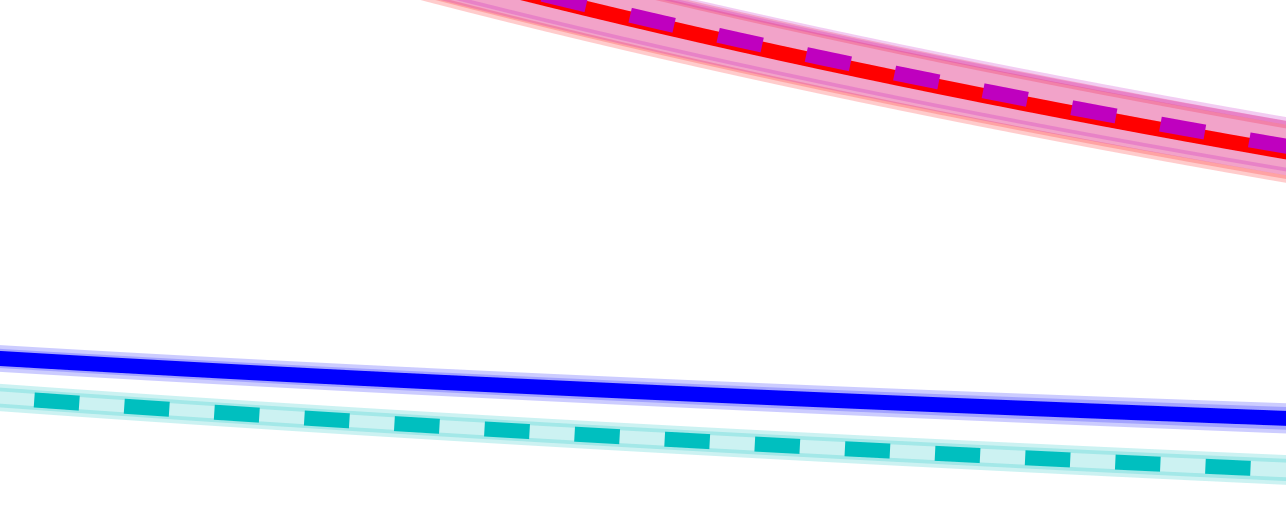}
    \includegraphics[width=0.45\textwidth]{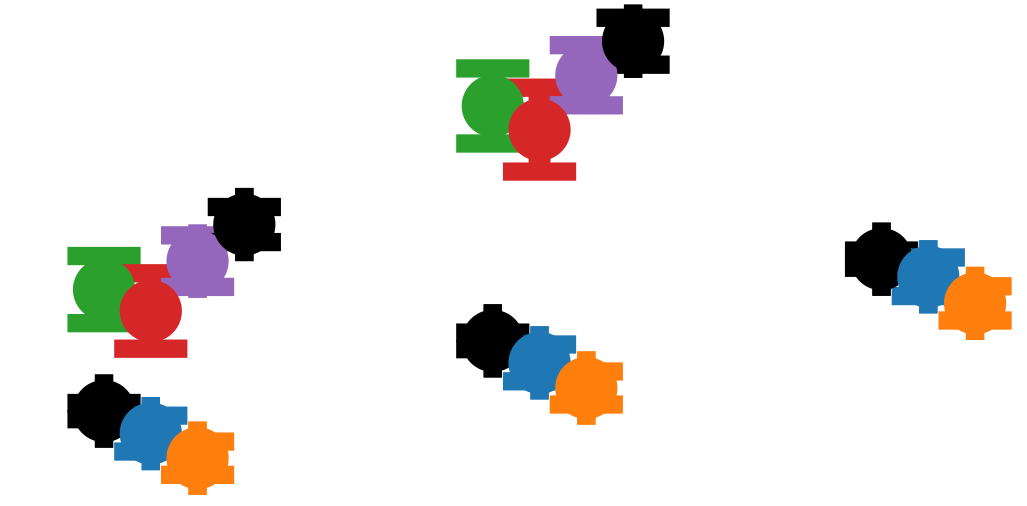}
    \caption{form factor results for $B_s \to K \ell \nu$ the RBC/UKQCD collaboration~\cite{Flynn:2023nhi} (left) and $D \to \pi \ell \nu$ by the Fermilab/MILC collaboration~\cite{FermilabLattice:2022gku} (right) as a function of the basis in which the chiral-continuum limit is performed. The bottom panels are zooms into regions of the top panels.}
    \label{fig:basischoices}
\end{figure}

In tree-level pseudoscalar to pseudoscalar decays often the $f_0$ form factor has the largest disagreement.
For the cases of $B_s \to K \ell \nu$ (and hence analogously $B \to \pi \ell \nu$) the RBC/UKQCD collaboration recently put forward a possible source of the discrepancy related to the form factor basis in which the chiral-continuum limit takes place~\cite{Flynn:2023nhi,Flynn:2023qmi}.
In particular they report a statistically significant difference depending on whether the chiral continuum extrapolation is performed in the HQET form-factor basis ($f_\parallel$ and $f_\perp$) or in the relativistic ($f_0$ and $f_+$) basis.
The difference is shown in the left panels of Figure~\ref{fig:basischoices} where with increasing final state momentum a discrepancy between the solid blue and the dashed cyan lines arises.
This discrepancy persists when considering a full error budget.
Whilst this predominantly affects $f_0$, the constraint imposing $f_+(q^2=0)=f_0(q^2=0)$ which is typically used in the kinematic extrapolation means that this still influences the form factor $f_+$ and hence also the value of $V_{ub}$ that will be extracted.
Similar statistically significant dependencies on the choice of extrapolation bases have been observed by the FNAL/MILC collaboration in their study of semileptonic decays of $D_{(s)}$ mesons~\cite{FermilabLattice:2022gku} as can be seen from the plot on the right hand side of Figure~\ref{fig:basischoices}. Whilst providing some indication of possible origins of some of the tensions in the literature, in order to fully understand and then resolve these, more independent calculations are urgently required. Scrutiny these results would be further simplified by being able to compare intermediate and less complicated quantities as will be discussed in Sec.~\ref{sec:benchmarks}.

\subsection{Recent and ongoing work}
Several collaborations are working on updates and/or new calculations that are aimed to address these issues. In particular the $\chi$QCD collaboration presented their progress on $D \to K \ell \nu$ decays~\cite{Shen:2025njo} using overlap fermions for all quarks on ensembles generated with domain wall fermions in the sea. Results are presented for a number of light quark masses, but currently limited to a single lattice spacing, so that no continuum limit can be taken yet.
The RBC/UKQCD~\cite{Boushmelev:2024jqg} provided a status report on the calculation of $B_s \to D_s^* \ell \nu$ form factors. These calculation also takes place on RBC/UKQCD's $N_f=2+1$ flavour ensembles with domain wall fermions being used for valence light, strange and charm~\cite{Boyle:2016imm} quarks, whilst the bottom quark is simulated using relativistic heavy quarks. Initial results are presented for a single ensemble but a larger dataset with additional lattice spacings and pion masses is currently being analysed.

In Ref.~\cite{Butti:2025rlu} the Fermilab/MILC collaborations presented the status of a follow-up calculation to their existing $B \to D^* \ell \nu$ work~\cite{FermilabLattice:2021cdg}. This work uses the highly improved staggered quarks (HISQ) $N_f=2+1+1$ ensembles for light, strange and charm quarks and the Fermilab action for the bottom quark. Preliminary blinded form factor results are shown for four lattice spacings and several pion masses, including the physical one. The chiral and continuum limit extrapolation have not been performed yet, but are currently ongoing work.

The HPQCD collaboration reported on work using HISQ quarks on $N_f=2+1+1$-HISQ ensembles for all valence quarks. In particular they compute $B,D \to \pi \ell \nu$ and $B_s,D_s\to K \ell \nu$ using three lattice spacings and pion masses down to the physical value~\cite{Roberts:2025bsp}. Preliminary form factor results are presented for multiple heavy-quark masses on a single ensemble. Analysing the remaining ensembles and the subsequent chiral-continuum-heavy-quark extrapolations are ongoing work. In Ref.~\cite{Harrison:2025kxm}, they furthermore report on an update of their previous $B_c \to J/\psi \ell \nu$ calculation~\cite{Harrison:2020gvo} additionally including an ensemble with physical pion masses and an ensemble which allows to directly simulate at the physical bottom-quark mass. A nice feature of this update is that the mass dependence on the susceptibilities (which enter the kinematic parameterisation) has been mapped out by the same collaboration and can hence be taken into account~\cite{Harrison:2024iad}.

Ref.~\cite{Meng:2024nyo} performed the first calculation of (vector to pseudoscalar) $J/\psi \to D_{(s)} \ell \nu$ decays using the CLQCD $N_f=2+1$ ensembles with Wilson-Clover fermions. This work is based on three gauge field ensembles with $M_\pi \sim 300\,\mathrm{MeV}$ but at different lattice spacings. Whilst for the $D_s$ final state the absence of physical pion masses or a chiral extrapolation is a pure sea-quark (and hence expected to be sub-leading) effect, this might not be the case for the final state $D$-meson. This investigation is deferred to future studies.

\subsection{Challenges in fully-relativistic calculations of semi-leptonic form factors containing bottom quarks}
The kinematically allowed range for these form factors depends on the masses of the initial and final states and covers $q^2 \in [0,(M_I-M_F)^2]$.
For initial states containing a bottom quark, its heavy mass sets a high scale, making it currently unfeasible to simulate the whole $q^2$-range at physical masses.
Furthermore the heavy-quark mass currently necessitates either the employment of an effective action for the heavy-quark mass~\cite{El-Khadra:1996wdx, Lin:2006ur, Christ:2006us, Lepage:1992tx}, or extrapolations from lighter-than-physical masses to the physical bottom-quark mass. 
In recent years great progress has been made and some first calculations with individual ensembles fine enough to directly simulate at (or very near) the physical $b$-quark mass have recently been presented~\cite{McLean:2019qcx,Cooper:2020wnj,Harrison:2020gvo,Harrison:2021tol,Cooper:2021bkt,Colquhoun:2022atw,Parrott:2022rgu,Harrison:2023dzh,Aoki:2023qpa}.
However in order to take the continuum limit the inclusion of unphysically-light quark-mass data points $m_h^{(i)} \ll m_b$ is currently still required, which in turn necessitates the additional limit $m^{(i)}_h \to m_b$ to be taken and controlled.
Of course, ultimately the fully-relativistic approach is systematically improvable and therefore favourable, whilst the effective action approaches are limited by some irreducible systematic uncertainties.
However, in the absence of calculations that take place entirely at physical quark masses, the two approaches are complementary as they require different extrapolations and have different sources of (dominant) systematic uncertainties.

Performing analyses with varying heavy-quark masses is challenging since this mass enters any given analysis via multiple and hence hard-to-disentangle effects. In particular\footnote{When simulations take place away from the physical pion mass, the first two statement are also true for final states that contain a valence light quark. However, this effect is typically milder, in particular when relating to the $q^2$ dependence, as the heavy-quark mass sets the scale that determines $q^2_\mathrm{max} = (M_I-M_F)^2$.}

\begin{itemize}
\item the form factors $f_X$ have a physical mass dependence on $m_h$.
\item the argument of the form factor $q^2$ has a physical mass dependence on $m_h$.
\item mass dependent discretisation effects of the form $(am_h)^n$ need to be parameterised.
\end{itemize}

For any given ensemble the heavy-quark-mass reach is limited by the lattice spacing, so that the available information for the continuum limit decreases as the quark mass is raised.
As a consequence, the parameterisation of the continuum limit typically relies heavily on the lighter-than-physical quark masses.
Similarly, the available coverage of $q^2$ depends strongly on the heavy-quark mass:
Whilst for moderate ``heavy''-quark masses around the charm scale it is possible to directly cover the full $q^2$ range, the size of the momenta that would need to be induced prohibits this when approaching the physical bottom-quark mass.
Hence, information about the low $q^2$-range is most strongly constrained by information obtained for far-from-physical heavy-quark masses.

As a result, current calculations necessarily rely strongly on data points far away from the physical value of the quark masses where the continuum limit is better controlled and the kinematically allowed range is significantly smaller. Additionally, since signal-to-noise properties tend to be more favourable at smaller heavy-quark masses, and different choices of the heavy-quark mass on a given ensemble are highly correlated, the data points furthest away from the desired physical kinematics carry the largest statistical weight in the fits.

\section{Suggested benchmark quantities for $b$-physics observables\label{sec:benchmarks}}
Since more and more analyses are using a fully-relativistic setup, they become increasingly more complicated and intricate\footnote{If at some point calculations only use points at physical quark masses, this will simplify remove necessary extrapolations and simplify the situation, but this will certainly still take some time.}. The multi-dimensional nature of the fit makes it more difficult to pin down the exact origin of any discrepancy between lattice results. In order to fully understand and scrutinise these results, particularly with view of existing and potential future tensions between lattice predictions, a number of benchmark quantities could be devised. In particular, going forward it would be very valuable for the community to provide some quantities that are well defined, can more easily be compared, and which either completely remove or significantly simplify some of the extrapolations, analogous to the ``window'' quantities for the hadronic vacuum polarisation contribution to the anomalous magnetic moment of the muon~\cite{RBC:2018dos}. A non-exhaustive list of suggestions for this was recently put forward in Ref.~\cite{Tsang:2023nay}.

\begin{description}
    \item[Separating the continuum limit from heavy quark and kinematic extrapolations:] Typically at least one form factor is well defined and accessible when initial and final state are at rest (i.e. the $q^2_\mathrm{max}$). Assuming that the (possibly unphysical) heavy quark masses have been tuned to the same value the continuum limit can be taken independently from any heavy quark or kinematic extrapolation. This also limits the degree to which a given continuum limit extrapolation relies on (sometimes hard to quantify) correlations to data at lighter ``far'' away heavy-quark masses with better statistical properties.
    
    \item[Separating the continuum limit from heavy quark extrapolations:]
    In cases of data points away from $q^2_\mathrm{max}$ the above is less trivial, since often the momentum units are prescribed by the accessible Fourier modes which depend on the (usually un-tuned) physical spatial extents of the ensembles. However, joint parametrisations of the kinematic dependence and the continuum limit can still take place at fixed heavy-quark mass. For parametrisations based on BGL-like $z$-expansions~\cite{Boyd:1994tt,Bourrely:2008za} this further avoids having to model the heavy quark mass dependence of pole masses or susceptibilities.
    
    The dependence on the kinematic extrapolation could be further reduced by either tuning the induced momenta (when using computational set-ups which are independent of the accessible Fourier modes), or by interpolating data to fixed reference kinematic values.
    
    \item[Agreeing on consensus intermediate heavy quark masses:] In principle the continuum extrapolated results at fixed heavy quark masses can be directly compared between collaborations, assuming that full uncertainty budgets are assembled.  It would be good if the community could agree on some ``canonical'' values (analogous to the prescription of isoQCD in $g-2$ caluclations). This would allow a direct comparison of intermediate results. This could also be used to perform sub-averages of the literature results. 

    A natural set of heavy-quark masses to be considered will include the physical charm- and bottom-quark masses. In addition, this could be complemented by multiples or fractions of the charm and bottom quark masses, or -- scheme independent and simpler to implement --  of $M_{D_{(s)}}$ and the $M_{B_{(s)}}$ masses, but for a convention to be maximally useful, it must necessarily be a community consensus.
    
    \item[Performing heavy-quark extrapolations at fixed kinematic choices:] When the heavy-quark mass varies so does the kinematic variable $q^2$. However, there are ways to define trajectories that allow for staying at ``fixed'' kinematics. The trivial cases are $q^2_\mathrm{max}$ and $q^2=0$, but fractions of $q^2_\mathrm{max}$ (or other kinematic variables) could be used to define a trajectory along which the heavy-quark mass extrapolation is performed.
    
    Furthermore, if quantities are designed in a way that allows a straightforward connection to the static limit, extrapolations can be turned into interpolations~\cite{Guazzini:2007ja,Sommer:2023gap}.

    Since these extrapolations are performed in the continuum limit, existing data from other calculations could be used to complement and hence improve upon the control over the heavy quark extrapolation.
\end{description}
Some of these quantities have already been produced in individual calculations, but a unified choice for some of the reference values would be highly advantageous. Of course, any final result quoted by a given calculation should be structured in whichever way the authors see fit, but additionally producing these benchmark quantities based on subsets of the data set and with full uncertainty budgets would enable us as a community to identify and react to tensions more quickly in a transparent process that will bolster confidence in the results within and outside the lattice community.

\section{Selection of novel results \label{sec:newresults}}
The following is an incomplete description of some of the works presented at this lattice conference and not yet covered in the above.
\subsection{$K \to \pi \pi$}
\begin{figure}
    \centering
    \includegraphics[width=0.55\linewidth]{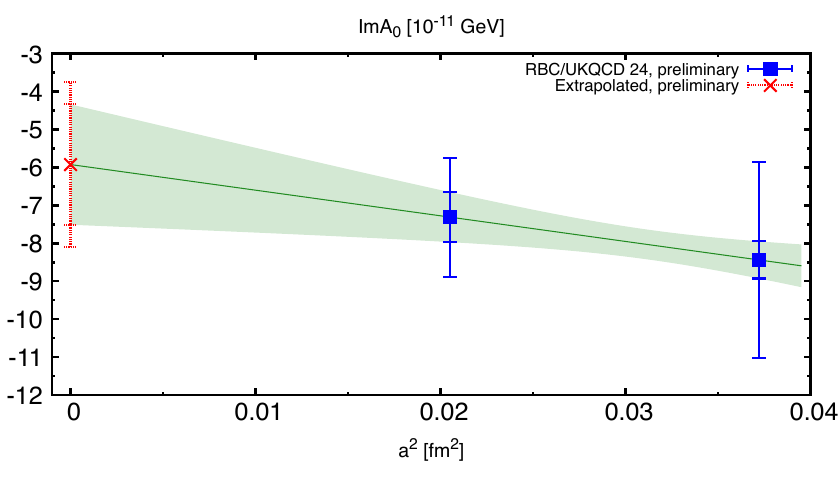}
    \includegraphics[width=0.43\linewidth]{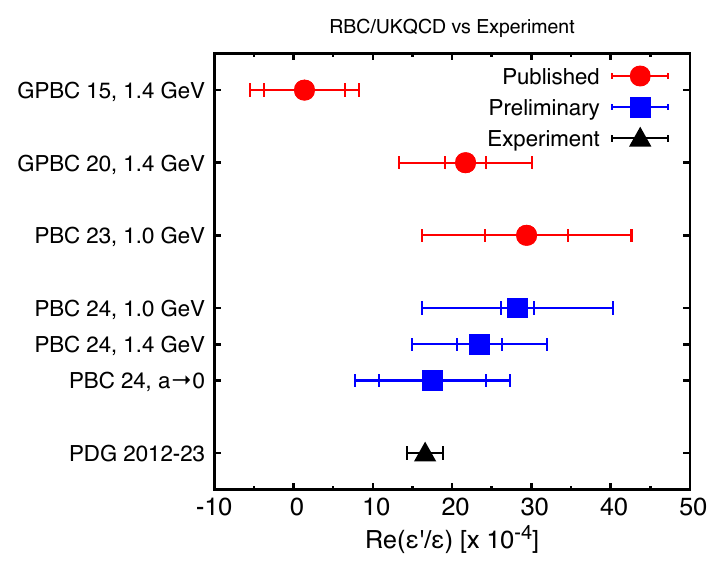}
    \caption{\emph{Left}: First continuum limit illustrated for the case of $\mathrm{Im}(A_0)$. \emph{Right}: Comparison of the literature values for the prediction of $\mathrm{Re}(\epsilon/\epsilon'$ to the PDG value. Plots are taken from Ref.~\cite{Tomii:2025lop}.}
    \label{fig:Kpipi}
\end{figure}
Masaaki Tomii presented an updated of the $K \to \pi \pi $ calculation pursued by the RBC/UKQCD collaboration~\cite{Tomii:2025lop}. This work extends the previous calculation~\cite{RBC:2023ynh} which took place at a single lattice spacing ($a^{-1} \approx 1.0\,\mathrm{GeV}$) by supplementing it with a second lattice spacing ($a^{-1}\approx 1.4\,\mathrm{GeV}$). The ensembles used have physical quark masses and use periodic boundary conditions (BCs) and are hence complementary to the existing calculation using G-parity BCs~\cite{RBC:2015gro,RBC:2020kdj}. The second lattice spacing allows - for the first time - to take a continuum limit which allows to quantify discretisation effects in a data-driven way (see left-hand panel of Figure~\ref{fig:Kpipi}). The authors then compute $\mathrm{Re}(\epsilon/\epsilon')$ and find good agreement with the literature and compatibility with the experimental value (right-hand panel). Due to the relatively coarse lattice spacings used in this study the authors are pursuing to supplement this work in the future with a significantly finer ($a^{-1} \approx 2.7\,\mathrm{GeV}$) lattice spacing. 

\subsection{Neutral meson mixing and life times}
\begin{figure}
    \centering
    \includegraphics[width=0.43\linewidth]{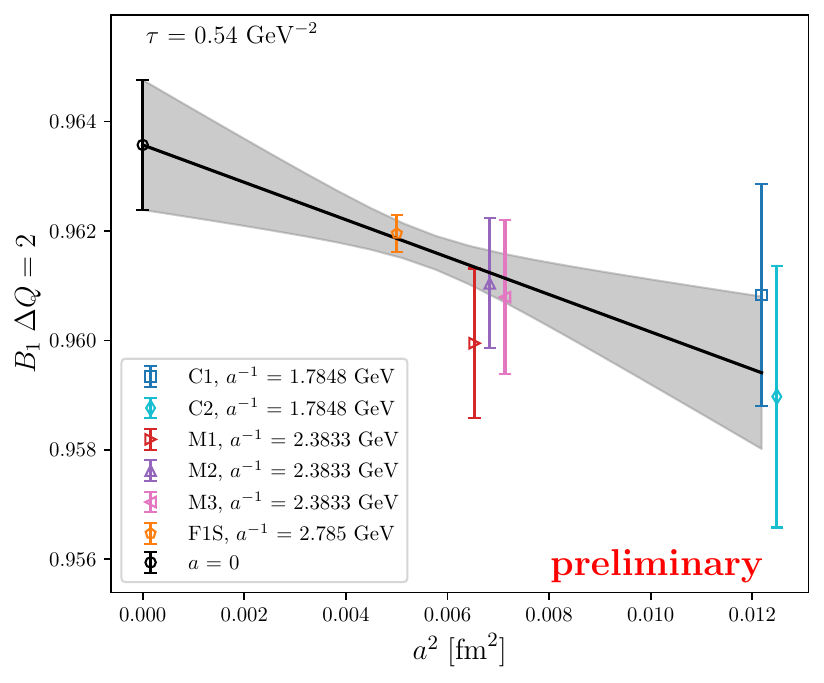}
    \includegraphics[width=0.56\linewidth]{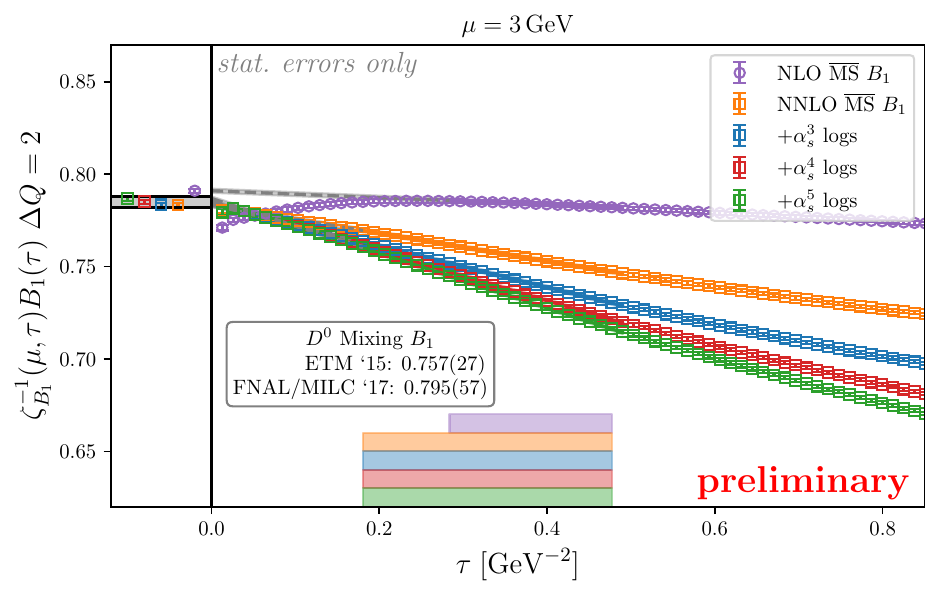}
    \caption{\emph{Left:} A sample continuum limit for the short distance contribution to neutral $D_{(s)}$ meson mixing illustrating the steps in the analysis. \emph{Right:} Continuum extrapolated results as a function of flow time, entering the small flow time expansion for different levels of perturbative matching. The coloured horizontal bands at the bottom of the plot indicate the ranges in $\tau$ used for the linear extrapolation to zero flow time. Plots are taken from Ref.~\cite{Black:2024iwb}.}
    \label{fig:GF}
\end{figure}
In Ref.~\cite{Black:2024iwb} Matthew Black presented progress in the calculation of neutral meson mixing parameters and life times using gradient flow~\cite{Narayanan:2006rf,Luscher:2010iy,Luscher:2013cpa} renormalisation and the short flow time expansion~\cite{Luscher:2011bx,Suzuki:2013gza,Luscher:2013vga}.
The idea is that gradient-flowing the relevant matrix elements removes ultraviolet divergences and hence can be used to renormalise these operators in a gradient flow scheme. The continuum limit can then be taken at finite flow time.
Using the short flow time expansion, these operators can be perturbatively related to the $\overline{\mathrm{MS}}$-scheme at finite flow time $\tau$ and the $\overline{\mathrm{MS}}$ result is recovered in the limit $\tau \to 0$.
This hence presents a window problem, since too large values of $\tau$ make the extrapolation for $\tau \to 0$ unreliable, whilst for too small values of $\tau$ the divergences have not been sufficiently removed.
As a proof of concept, they compute the $O_1$ operator for $\Delta Q = 2$ and $\Delta Q = 0$ for the (fictitious) neutral $D_s$-meson.
They use six of the $N_f=2+1$ RBC/UKQCD ensembles with domain wall fermions, comprising three lattice spacings and pion masses down to $\sim 270\,\mathrm{MeV}$.
The light and strange quark masses use the unitary value of the action, whilst the charm quark action is adapted for a better heavy-quark mass reach~\cite{Boyle:2018knm}.
For each value of the flow time the continuum limit needs to be taken -- one such continuum limit is illustrated in the left-hand panel of Figure~\ref{fig:GF}.
In a second step the result of this is matched to $\overline{\mathrm{MS}}$.
Different orders to which this matching is performed and considering additional logarithmic corrections lead to the differently coloured curves in the right-hand panel of the figure.
Finally a linear region in $\tau$ needs to be identified and the $\tau \to 0$ extrapolation performed.
The extrapolation ranges and extrapolated results can be seen at the bottom of the figure and to the left of the $\tau=0$ axis, respectively.
These initial preliminary results for the short distance contribution of $\Delta Q = 2$ for $B_1$ look promising and agree with the literature and (once finalised) will provide a proof of principle.
The preliminary results for $\Delta Q = 0$ are protected from mixing with lower dimensional operators but not from power divergences~\cite{Kim:2021qae} and are hence slightly more involved to finalise to the point that they can be phenomenologically interpreted, but the continuum limit and small flow time extrapolations look qualitatively similar to the $\Delta Q = 2$ case.

\begin{figure}
    \centering
    \includegraphics[width=0.32\linewidth]{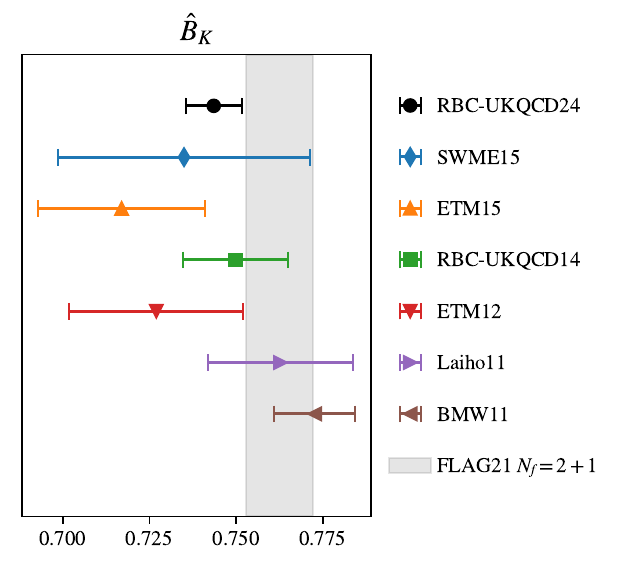}
    \includegraphics[width=0.6\linewidth]{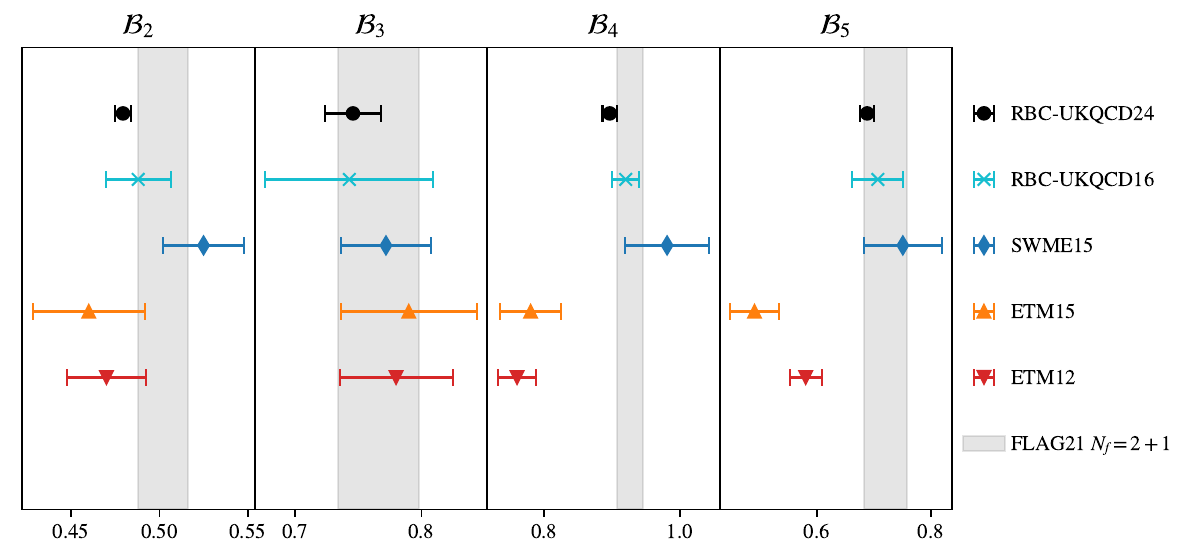}
    \caption{Literature comparison of the kaon bag parameter $\hat{B}_K$ (left) as well as the beyond-the-standard model bag parameters (right). Plots are taken from Ref.~\cite{Boyle:2024gge}.}
    \label{fig:KKbar}
\end{figure}
The RBC/UKQCD collaborations recently presented an update of the neutral kaon mixing bag parameters in and beyond the Standard Model~\cite{Boyle:2024gge}. This study uses $N_f=2+1$ domain wall fermion ensembles with two ensembles at the physical pion mass and three lattice spacings. The non-perturbatively computed renormalisation factors are extracted the massless limit. The results are the most precise to date and are in good agreement with previous calculations using non-exceptional renormalisation schemes as can be seen in Figure~\ref{fig:KKbar}. Work to extend this to neutral meson mixing in the $B_{(s)}$-sector remains ongoing~\cite{Boyle:2021kqn}. It is worth noticing that the leading uncertainty in this work arose from the perturbative matching to $\overline{\mathrm{MS}}$. This situation has recently been improved by the first two-loop matching calculation~\cite{Gorbahn:2024qpe} between RI-(S)MOM and $\overline{\mathrm{MS}}$.

\subsection{Results for $\epsilon_K$ using lattice QCD inputs}
The SWME collaboration presented their recent analysis for $\epsilon_K$ based on lattice inputs~\cite{Jwa:2024xjq,Jwa:2025fon}. Combining world averages for lattice inputs such as $\hat{B}_K$, $\xi_0$, $\xi_2$, $\xi_{LD}$, $m_c(m_c)$, $F_K$, $|V_{ud}|$, $|V_{us}|$ and $|V_{cb}|$ they compute the Standard Model value of $(\epsilon_K)_\mathrm{SM}$ and compare it against the experimentally measured value $(\epsilon_K)_\mathrm{exp}$. They report tensions of $\sim 5 \sigma$ between $|\epsilon_K|_\mathrm{SM}$ and $|\epsilon_K|_\mathrm{exp}$ when using values for $|V_{cb}|$ stemming from exclusive determinations but find that these tensions disappear when instead using $|V_{cb}|$ from inclusive determinations (which are however currently not based on lattice QCD calculations).

Whilst at present form factors that give access to $|V_{cb}|$ from exclusive decays do not seem display any major tensions, the number of results is limited and further calculations are required to bolster confidence in the exclusive determinations of $V_{cb}$. Furthermore, tensions for the extraction of $V_{cb}$ are observed between different combinations of lattice form factors~\cite{FermilabLattice:2021cdg, Harrison:2023dzh, Aoki:2023qpa} and experimental datasets as discussed in Refs.~\cite{Martinelli:2022xir,Fedele:2023ewe,Bordone:2024weh} as well as in some additional observables such as forward-backward asymmetries. As discussed above, several form factor calculations are currently actively pursued and will provide further insight into this in the near future, whilst experimental measurements are also expected to improve.

\begin{figure}
    \centering
    \includegraphics[width=.45\linewidth]{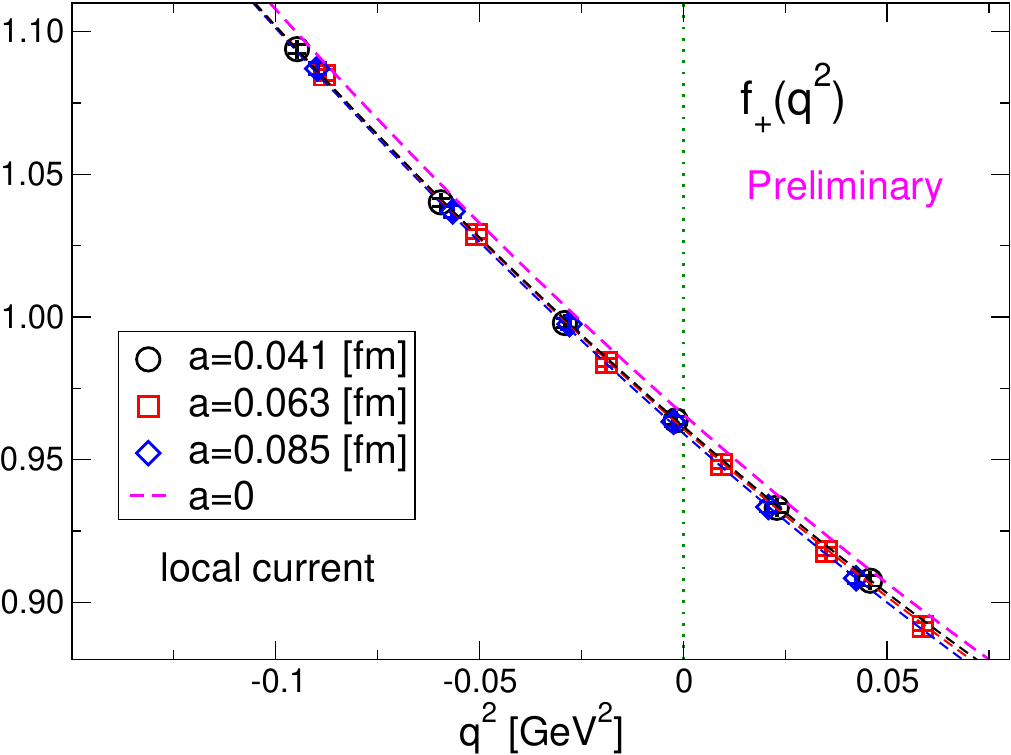}
    \includegraphics[width=.45\linewidth]{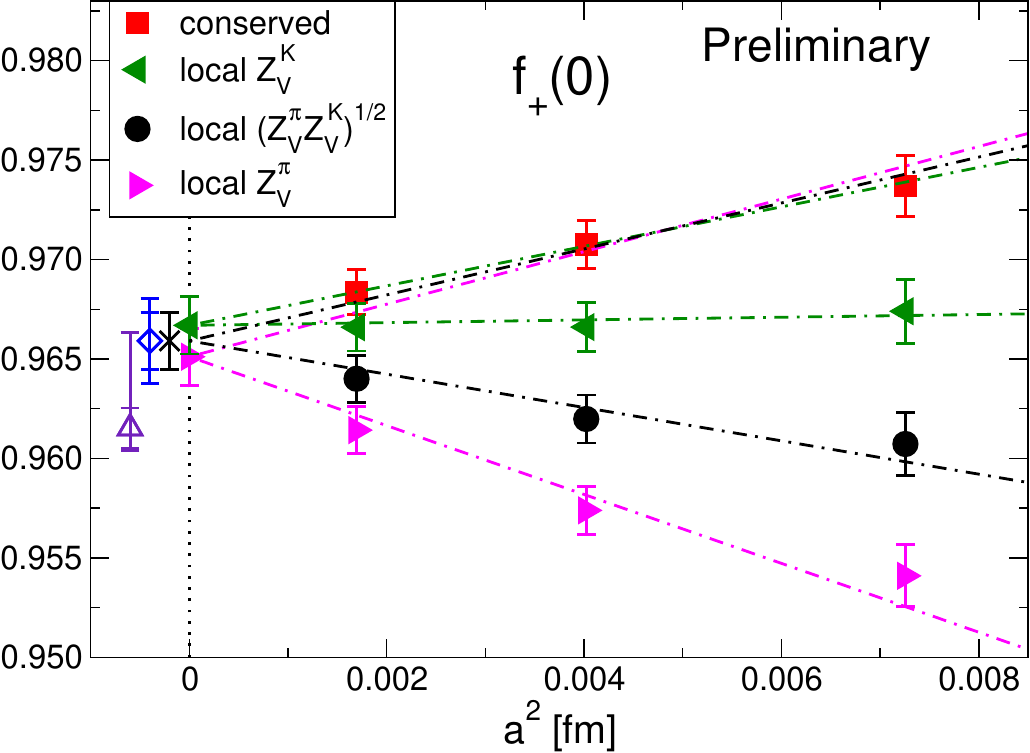}
    \includegraphics[width=.6\linewidth]{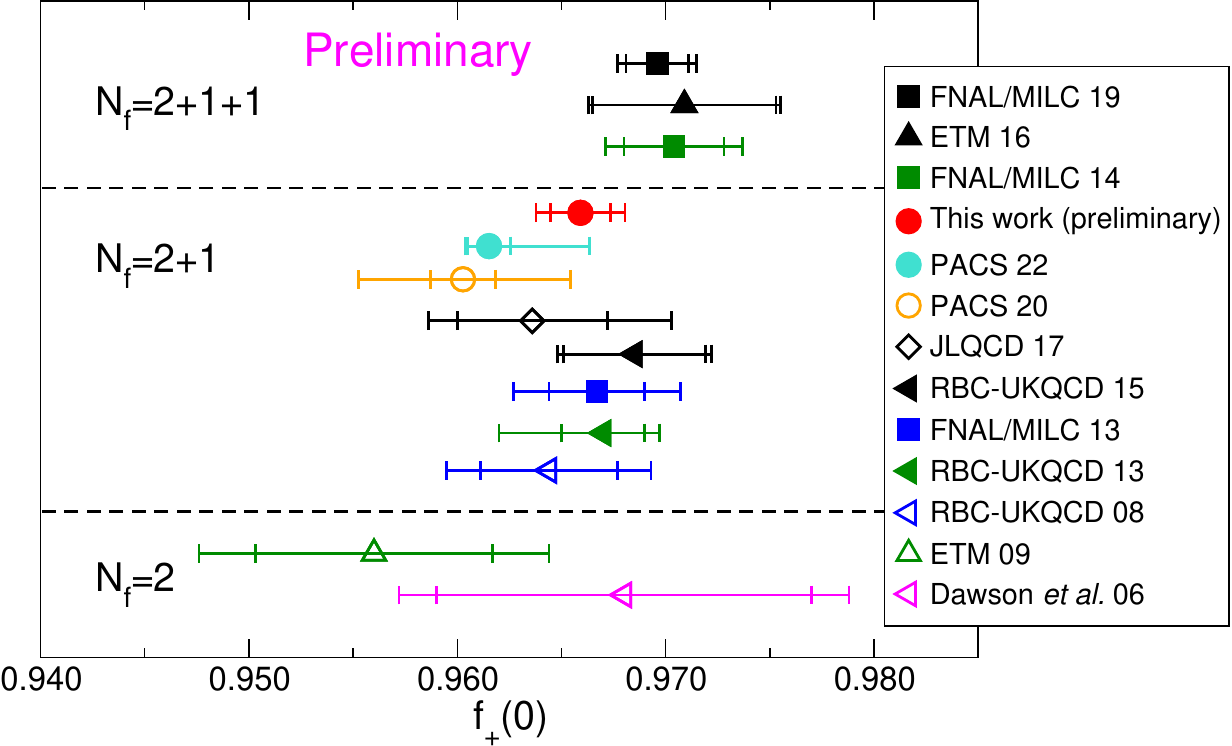}
    \caption{Continuum limit of the $K_{\ell 3}$ form factor $f_+$ a function of $q^2$ (top left). The continuum limit of $f_+(0)$ as a function of different renormalisation choices (top right) and the comparison of $f_+(0)$ to values in the literature. Plots taken from Ref.~\cite{Yamazaki:2024otm}. }
    \label{fig:PACSKl3}
\end{figure}

\subsection{$K_{\ell3}$} The PACS collaboration presented an update of $K_{\ell 3}$ form factors using large volume $(10\,\mathrm{fm})^4$ ensembles at near-physical quark masses and three lattice spacings~\cite{Yamazaki:2024otm}. The top left panel of Figure~\ref{fig:PACSKl3} shows the $f_+$ form factor as a function of the momentum transfer $q^2$ for the three ensembles as well as the continuum limit extrapolated result. The top right panel investigates different definitions of the renormalisation constants which all nicely agree in the continuum limit. Whilst these results are still preliminary, they observe nice agreement with the existing literature (bottom panel) with competitive uncertainties. 

\subsection{$B_s \to \mu\mu\gamma$}
\begin{figure}
    \centering
        \includegraphics[width=0.35\linewidth]{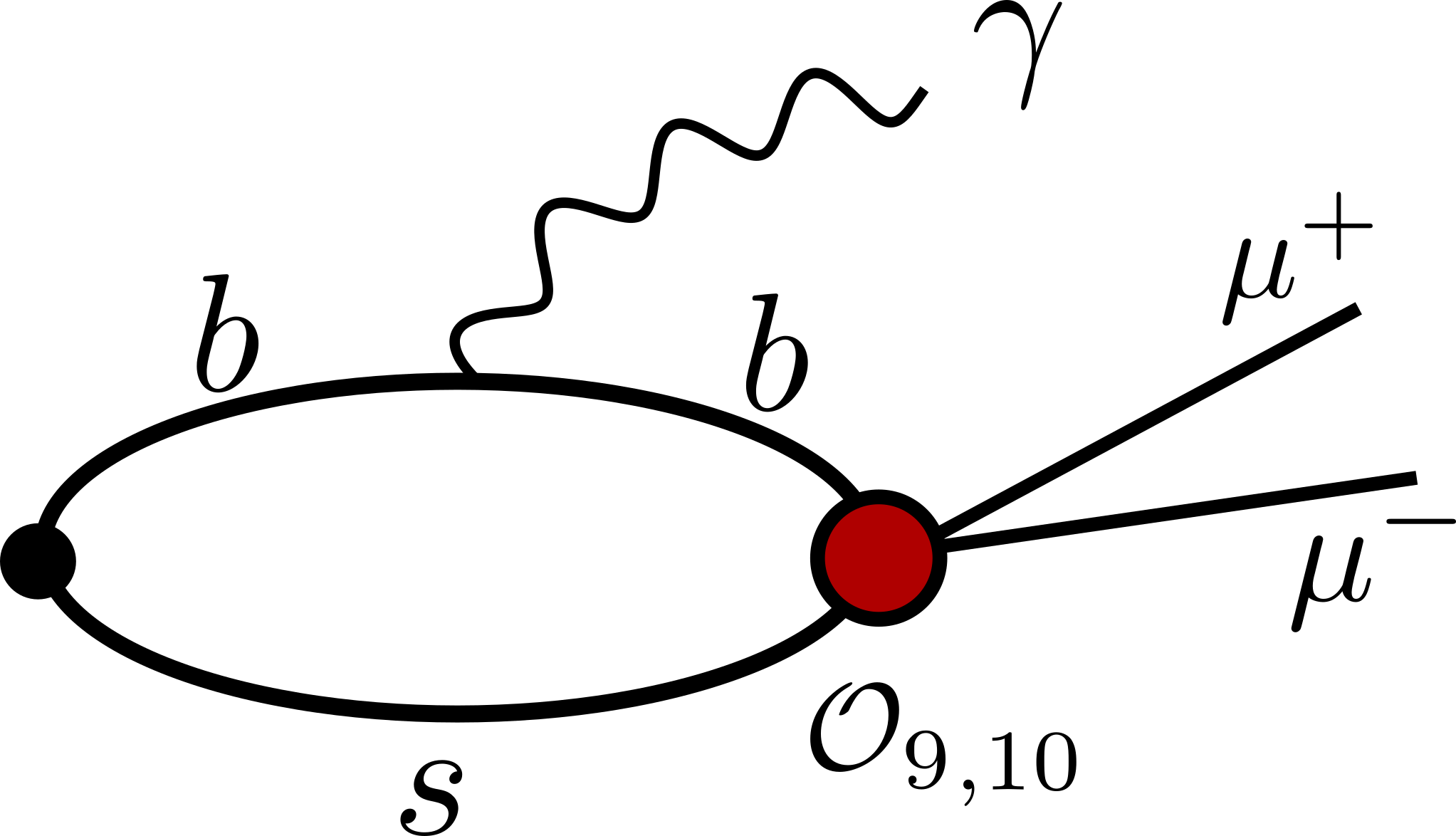} 
        \hspace{1cm}
        \includegraphics[width=0.40\linewidth]{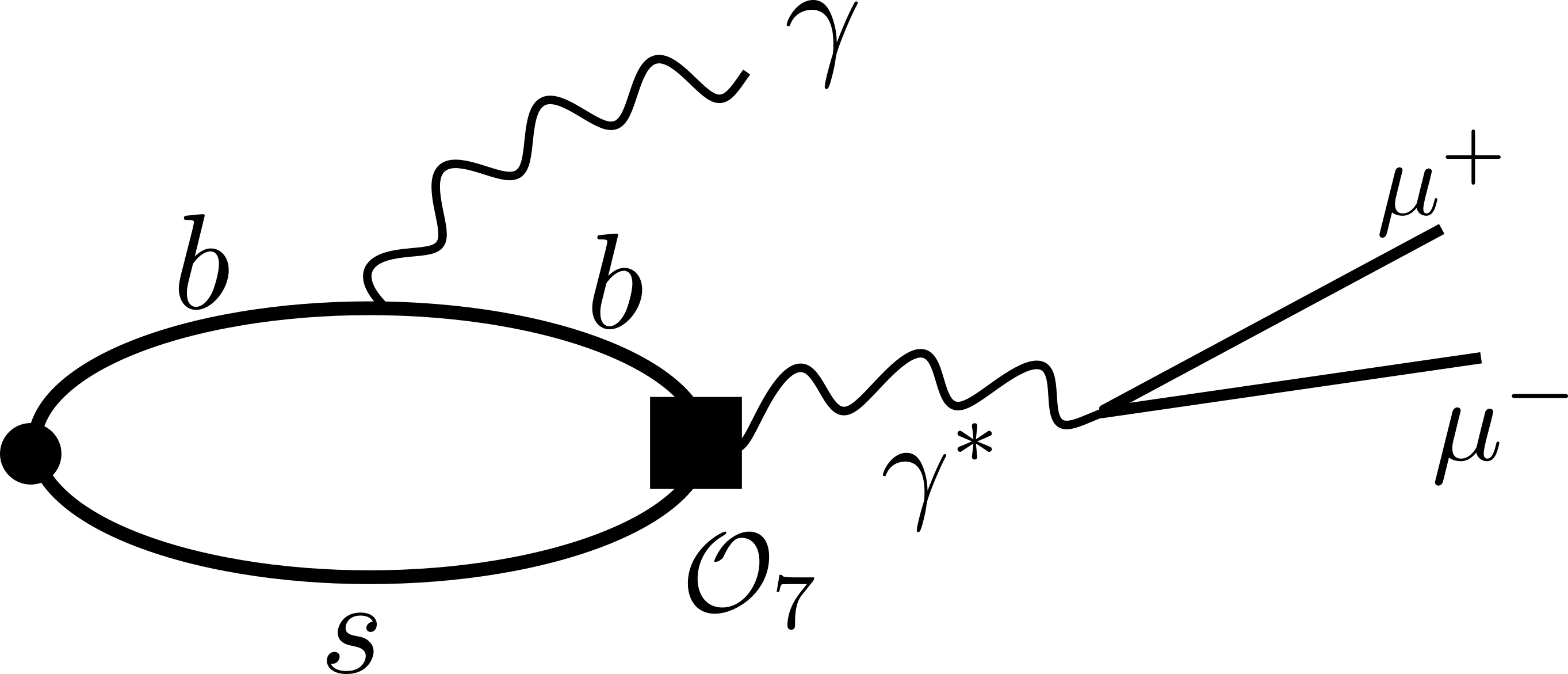}       
    \includegraphics[width=1\linewidth]{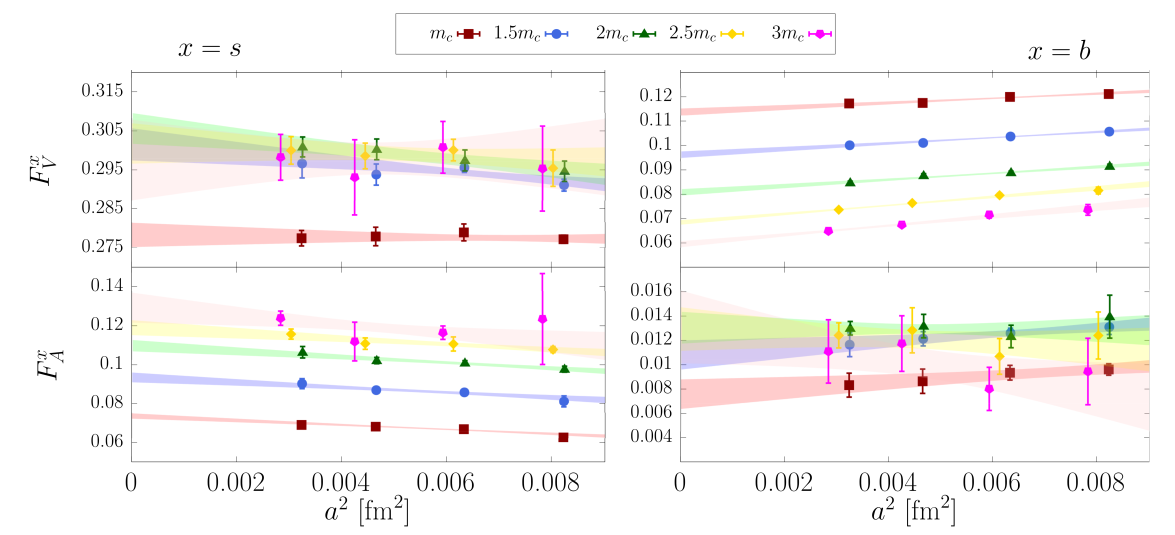}
    \caption{\emph{Top}: Relevant quark flow diagrams for $F^b_A$, $F^b_T$, $F^b_{TV}$ and $F^b_{TA}$, where the superscript $b$ denotes the fact that the photon is emitted from the b-quark. The equivalent diagrams with $b \leftrightarrow s$ are also computed and lead to the corresponding form factors with a superscript $s$. \emph{Bottom}: Sample continuum limits for $F^x_A$ and $F_V^x$ for $x=s,b$ at a fixed $x_\gamma=0.4$. Plots taken from Refs.~\cite{Frezzotti:2024kqk,Frezzotti:2024chv}.}
    \label{fig:RmSoton-CLs}
\end{figure}
\begin{figure}
    \centering
    \includegraphics[width=\linewidth]{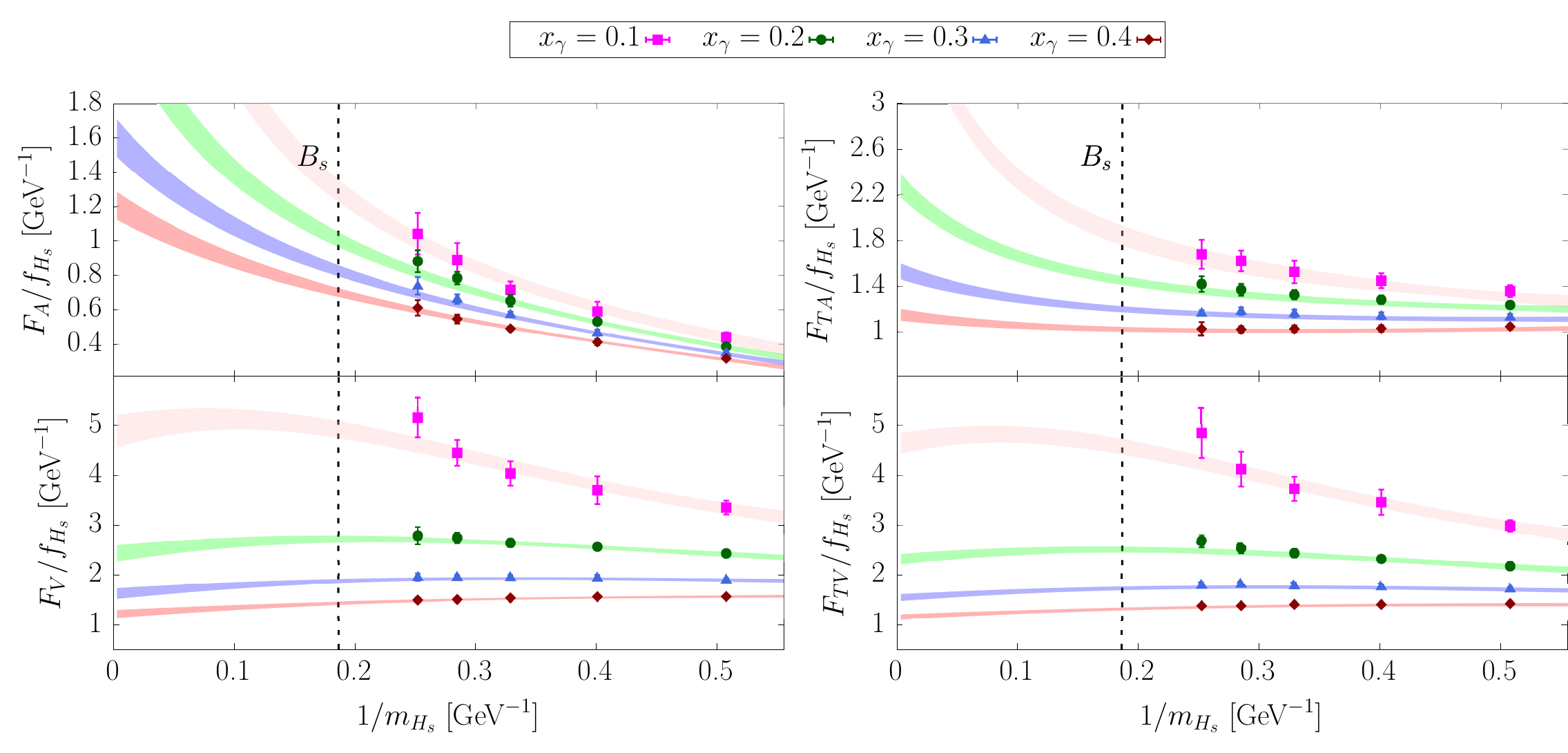}
    \caption{Extrapolation of the form factors in the continuum limit to the physical bottom quark mass. Plots taken from Ref.~\cite{Frezzotti:2024kqk}.}
    \label{fig:RmSoton-hq}
\end{figure}
\begin{figure}
        \centering
        \includegraphics[width=0.32\linewidth]{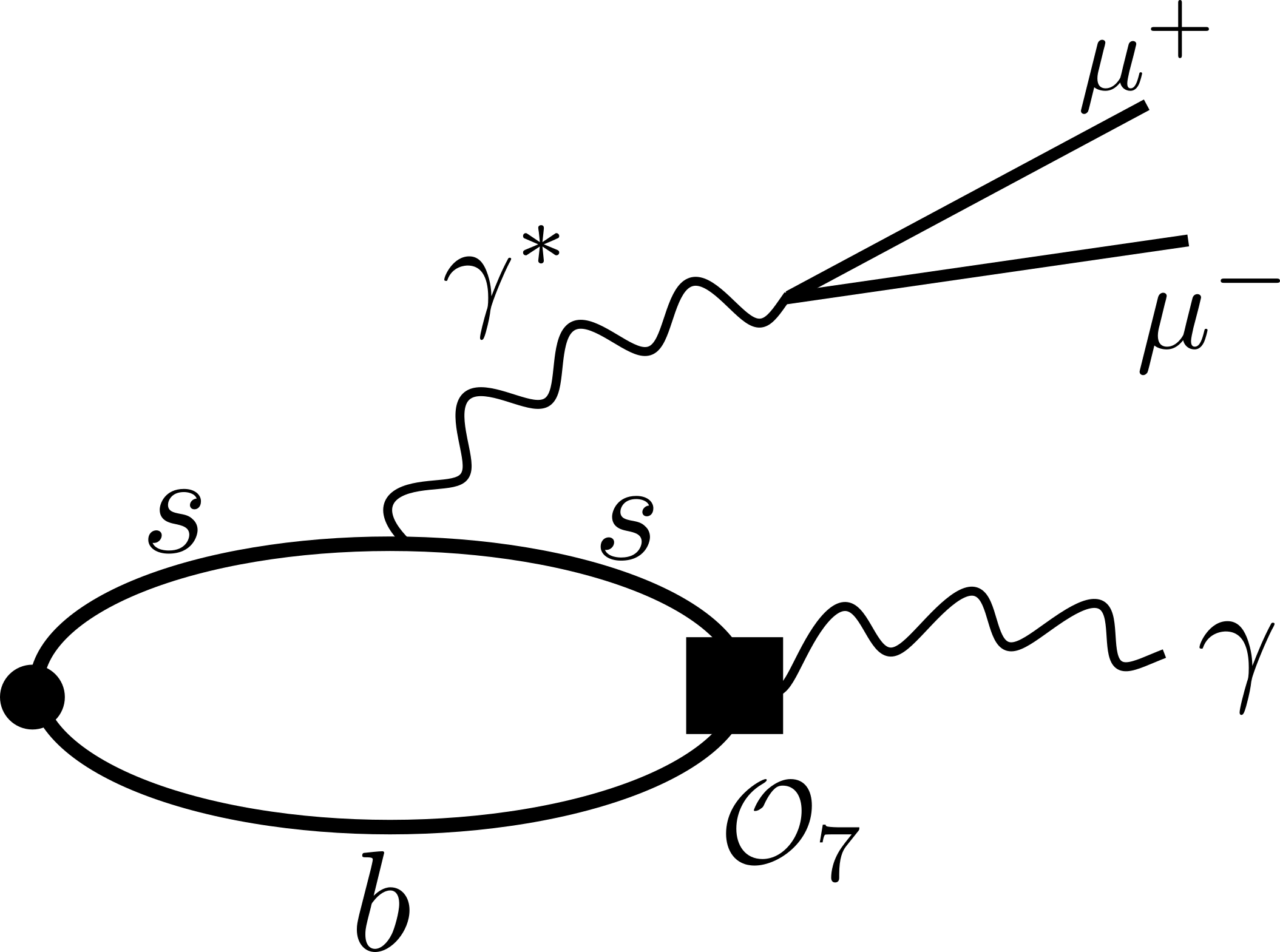}
        \hspace{1cm}
        \includegraphics[width=0.47\linewidth]{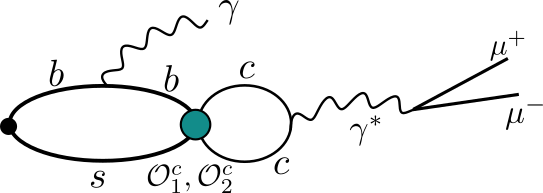}
        \caption{Diagrams relating to the local form factor $\overline{F}_T$ (left) and non-local charm loops (right). Plots taken from Ref.~\cite{Frezzotti:2024kqk}.}
        \label{fig:RmSoton_rest}
    \end{figure}

Francesco Sanfilippo presented the recent work of the Rome-Southampton collaborations on the $B_s \to \mu\mu\gamma$ decay rate at large $q^2$~\cite{Frezzotti:2024kqk}.
Since the helicity suppression is lifted, this is expected to be comparable in size to $B_s \to \mu\mu$.
The calculation uses the ETMC $N_f=2+1+1$ Wilson-Clover twisted mass ensembles at four lattice spacings and including physical pion mass ensembles.
Five heavy-quark masses are simulated corresponding to heavy-strange meson masses in the range $M_{H_s}\in [M_{D_s},2M_{D_s}]$.
In the decomposition, five local form factors arise which are computed from first principles.
These form factors depend on the kinematic variable $x_\gamma \equiv 1-q^2/M_{hs}^2$ for which four values are calculated and where $q^2$ is the momentum transfer to the di-muon pair. The four dominant factors arise from the diagrams depicted in the top of Figure~\ref{fig:RmSoton-CLs} and are extracted from suitably defined ratios of correlation functions. Subsequently, each of these four form factors is extrapolated to the continuum for each choice of the heavy quark mass, each choice of $x_\gamma$ and for both choices of which quark is emitting the photon, resulting in a total of 160 continuum limits. This is illustrated in the bottom of Fig.~\ref{fig:RmSoton-CLs} for two of the form factors and $x_\gamma = 0.4$. 
Having obtained the continuum limits, it remains to extrapolate these form factors to the physical bottom-quark mass. This extrapolation is guided by HQET scaling laws~\cite{Beneke:2011nf,Beneke:2020fot} and illustrated in Figure~\ref{fig:RmSoton-hq}.
The remaining local form factor can be obtained from the diagram depicted in the left-hand side of Fig.~\ref{fig:RmSoton_rest}. However, the hadronic tensor that needs to be computed cannot be analytically continued to Euclidean time for the simulated values of $x_\gamma$. This can be circumvented by using recently suggested inverse problem techniques~\cite{Frezzotti:2023nun,Hansen:2019idp}. Omitting the details since inverse problems are covered in Ref.~\cite{Jay:2025dzl}, they obtain a continuum and heavy-quark extrapolated result for $\overline{F}_T$ which turns out to be strongly suppressed. The final result of this impressive calculation is a prediction for the local form factors for large $q^2>(4.16\,\mathrm{GeV})^2$. 

Finally, taking the non-local contribution arising from the "charming penguins" (see right hand diagram in Fig.~\ref{fig:RmSoton_rest}) into account as a phenomenological shift of the corresponding Wilson coefficient $C_9$~\cite{Kruger:1996cv,Guadagnoli:2023zym,Kozachuk:2017mdk}, these form factors can be used to compute the branching fraction $\mathcal{B}(B_s \to \mu^+\mu^-\gamma)$ for the upper-cut presented in the bound set by the LHCb collaboration~\cite{LHCb:2024uff} and can hence be directly compared to these bounds. Whilst currently the experimental bound is above the theory prediction, efforts to improve on these measurements are ongoing. 

\subsection{Further work not covered here}
Due to time and space constraints, and the breadth of activity in this field, I was not able to cover every single contribution to this conference in detail. I refer the interested reader to the corresponding proceedings or recent publications~\cite{DeSantis:2025lal, Hu:2025hpn, Kellermann:2024jqg, ExtendedTwistedMass:2024myu, Chao:2024eil, Chao:2024vvl, DiCarlo:2025uyj, Ma:2023kfr, Tuo:2024bhm,Hodgson:2025iit} as well as to the plenary proceedings by Will Jay~\cite{Jay:2025dzl} and Felix Erben~\cite{Erben:2025zph}.

\section{Progress in methodology \label{sec:methods}}
Some of the most challenging parts of flavour physics precision calculations are 
\begin{itemize}
    \item the correct identification and extraction of ground state parameters
    \item the continuum extrapolation
    \item the extrapolation to the physical bottom quark mass (where required)    
\end{itemize}
All three of these areas have seen recent methodological improvements with new methods being suggested and/or implemented. Since the third point has already been included in the suggestion of the benchmarks above, I will now focus on the progress on the remaining two items.

\subsection{Excited state contamination}
Based on heavy meson chiral perturbation theory, concerns about enhanced excited state contaminations due to $B^* \pi$ states have recently been raised~\cite{Bar:2023sef}. These are expected to be small corrections for the two point functions, but particularly enhance the $f_\perp$ form factor in/near the static limit, raising concerns about how well these effects can be captured in multi-state fits. At this conference Antoine G\'erardin presented an investigation of these for the ALPHA collaboration employing different types of smearing (Gaussian smearing and distillation profiles). The outcome of this study will be important for any precision calculation of semi-leptonic form factors, in particular when (near) physical pion masses are used.

\subsection{Methodological improvements of the continuum limit}
The continuum limit is one of the most critical extrapolations to be taken in precision lattice simulations and can often be one of the leading sources of uncertainty.
Improving the continuum limit has the potential to make these extrapolations more reliable and hence the calculations more accurate.
Whilst not constraint to heavy-quark physics, they often introduce especially large discretisation effects.
In Ref.~\cite{Boyle:2016wis} a massive momentum subtraction scheme was suggested which has the potential to absorb large discretisation effects into renormalisation constants, resulting in a better behaved continuum limit without compromising the continuum Ward Identities and which smoothly connects to the massless limit. In this prescription the renormalisation conditions are imposed at some mass scale $\overline{m}$ which is a free parameter (to the extent that the required vertex function can be computed on the ensembles under consideration). The hope now is that appropriately choosing this parameter for a given quantity provides a \emph{tunable} parameter that will allow a flat approach to the continuum limit. 

\begin{figure}
    \centering
    \includegraphics[width=0.4\linewidth]{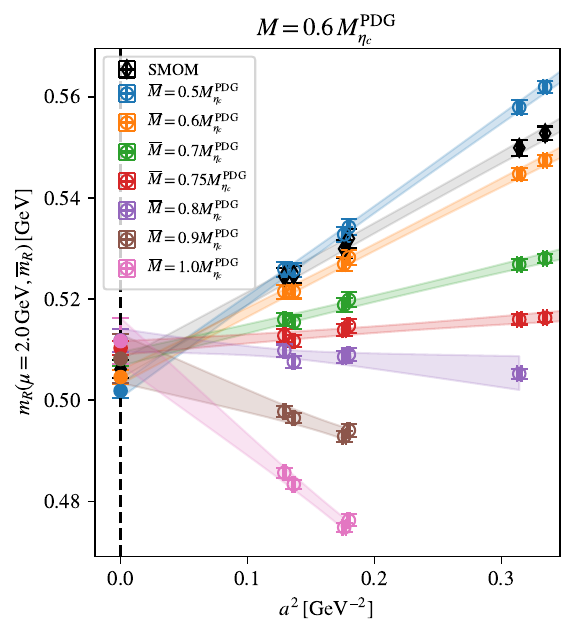}
    \includegraphics[width=0.59\linewidth]{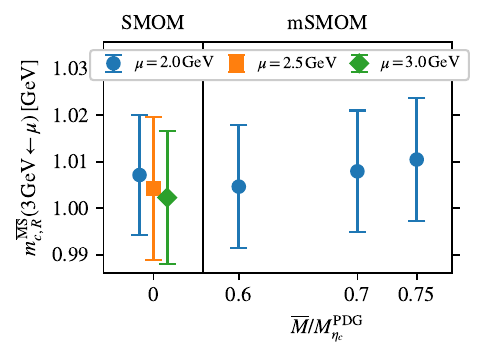}  
    \caption{\emph{Left}: Different approaches to the continuum limit in the massive SMOM scheme as a function of the mass defining the renormalisation conditions. \emph{Right}: comparison of results for the charm quark mass converted to $\overline{\mathrm{MS}}(3\,\mathrm{GeV})$ but obtained from different choices for the renormalisation scale $\mu$ and the quark mass $\overline{m}$ at which the renormalisation conditions are imposed. Plots taken from Ref.~\cite{DelDebbio:2024hca}}
    \label{fig:mSMOM}
\end{figure}

The RBC/UKQD collaboration presented a first numerical application of these schemes, demonstrating that this is indeed the case and used it to make a prediction for the charm quark mass~\cite{DelDebbio:2024hca}. The left-hand plot of Figure~\ref{fig:mSMOM} demonstrates that such an improvement in the continuum limit can indeed be achieved, whilst the right-hand plot demonstrates that the result once converted into a common scheme is independent of this choice. This encouraging result might be applicable to several other quantities that require renormalisation and therefore allow for better controlled continuum limits for a number of observables.

\section{Conclusions and outlook \label{sec:outlook}}
Flavour physics remains an active and exciting field of research in lattice
QCD. Whilst a number of tensions in ``known'' observables persist and will need
to be addressed, several calculations are ongoing which aim to achieve this. At
the same time the community is pursing the calculation of novel observables. Work on aspects relating to all sources of systematic uncertainties is also carried out, continuously improving the precision and the reliability of results. 
\section*{Acknowledgements}
I am grateful to the organisers for the invitation to give this talk as well as for kindly accommodating stringent timetabling constraints on my part. I would further like to thank all those who sent me material or made me aware of recent results prior to the conference. Finally, I would like to thank the CERN theory group, all my mentors and collaborators for their support in research and beyond.

\bibliographystyle{JHEP}
\bibliography{biblio}



\end{document}